\documentclass[a4paper,11pt]{article}
\usepackage{jheppub} 

\title{\boldmath Phase structure and critical phenomena in 2-flavor QCD by holography }

\author[a]{Yan-Qing Zhao,}
\author[b,c]{Song He,}
\author[a]{Defu Hou,}
\author[d,e,f]{Li Li,}
\author[g]{Zhibin Li}
\affiliation[a]{Institute of Particle Physics and Key Laboratory of Quark and Lepton Physics (MOS),\\
Central China Normal University,\\Wuhan 430079, China}
\affiliation[b]{Center for Theoretical Physics and College of Physics, Jilin University,\\Changchun 130012, China}
\affiliation[c]{Max Planck Institute for Gravitational Physics (Albert Einstein Institute),\\Am Muhlenberg 1, 14476 Golm, Germany}
\affiliation[d]{CAS Key Laboratory of Theoretical Physics, Institute of Theoretical Physics,\\
	Chinese Academy of Sciences,\\Beijing 100190, China}
\affiliation[e]{School of Fundamental Physics and Mathematical Sciences,\\
	Hangzhou Institute for Advanced Study, University of Chinese Academy of Sciences,\\Hangzhou 310024, China}
\affiliation[f]{Peng Huanwu Collaborative Center for Research and Education, Beihang University,\\Beijing 100191, China}
\affiliation[g]{School of Physics and Microelectronics, Zhengzhou University,\\Zhengzhou 450001, China}
\emailAdd{zhaoyanqing@mails.ccnu.edu.cn}
\emailAdd{hesong@jlu.edu.cn}
\emailAdd{houdf@mail.ccnu.edu.cn}
\emailAdd{liliphy@itp.ac.cn}
\emailAdd{lizhibin@zzu.edu.cn}

\abstract{We explore the phase structure of Quantum Chromodynamics (QCD) with two dynamical quark flavors at finite temperature and baryon chemical potential, employing the non-perturbative gauge/gravity duality approach. Our gravitational model is tailored to align with state-of-the-art lattice data regarding the thermal properties of multi-flavor QCD. Following a rigorous parameter calibration to match equations of state and the QCD trace anomaly at zero chemical potential derived from cutting-edge lattice QCD simulations, we investigate thermodynamic quantities and order parameters. We predict the location of the critical endpoint (CEP) at $(\mu_{\text{CEP}}, T_{\text{CEP}})=(219,182)$ MeV at which a line of first-order phase transitions terminate. We compute critical exponents associated with the CEP and find that they almost coincide with the  critical exponents of the quantum 3D Ising model.}

\begin{document}
\maketitle
\flushbottom

\section{Introduction}\label{sec:00_intro}
A thorough understanding of the Quantum Chromodynamics (QCD) phase structure at specific temperature and density regimes is not only essential for elucidating the formation of matter but also for interpreting and predicting the wealth of data amassed from ongoing and future experiments involving heavy-ion collisions. While significant progress has been made in elucidating the phase structure at lower densities using cutting-edge lattice technology in recent years, challenges persist at higher densities, including the well-known sign problem~\cite{deForcrand:2009zkb}. Therefore, a robust, non-perturbative method is paramount at this juncture.

Numerous effective low-energy models have been developed to explore the Quantum Chromodynamics (QCD) phase diagram under various non-perturbative conditions. These include the Dyson-Schwinger equations (DSE)~\cite{Gao:2016qkh, Qin:2010nq, Shi:2014zpa, Fischer:2014ata}, the Nambu-Jona-Lasinio (NJL) model~\cite{Schwarz:1999dj, Zhuang:2000ub, Qiu:2023ezo, Qiu:2023kwv}, the Polyakov-Nambu-Jona-Lasinio (PNJL) model~\cite{Li:2018ygx, McLerran:2008ua, Sasaki:2010jz, Sun:2023dwh}, the functional renormalization group (fRG)~\cite{Fu:2019hdw, Zhang:2017icm}, hadron resonance gas models~\cite{Fujimoto:2021xix, Becattini:2016xct}, the coalescence model~\cite{Sun:2018jhg}, and a combination of DSE and fRG~\cite{Gao:2020qsj}. Some of these models predict the existence of a critical endpoint (CEP) where the first-order phase transition line terminates and transitions into a smooth crossover at small chemical potentials $\mu_B$. These predictions align well with results from lattice simulations~\cite{Burger:2014xga, Gavai:2008zr, Pisarski:1983ms, Aoki:2006we, Borsanyi:2010bp, Gottlieb:1992ii}.

An increasingly popular non-perturbative approach for studying Quantum Chromodynamics (QCD) involves the application of gauge/gravity duality~\cite{Maldacena:1997re, Gubser:1998bc, Witten:1998qj, Witten:1998zw} to construct holographic QCD models that describe QCD matter. This is achieved through both top-down~\cite{Babington:2003vm, Kruczenski:2003uq, Sakai:2004cn, Sakai:2005yt, Misra:2019thm} and bottom-up~\cite{Gubser:2008yx, Gursoy:2008bu} approaches. Notably, within the bottom-up framework, the Einstein-Maxwell-Dilaton (EMD) gravity model has been widely employed to create holographic QCD models that align with state-of-the-art lattice QCD simulations. Two common methods have emerged. The first one is the potential reconstruction method~\cite{Cai:2012xh,Li:2011hp,Li:2012ay,Cai:2012eh}, with recent developments discussed in~\cite{He:2020fdi, Arefeva:2018hyo, Bohra:2019ebj, Bohra:2020qom}. 
A limitation of this approach lies in its inability to quantitatively capture the thermodynamic behavior of lattice QCD simulations, suggesting potential improvement via better function configurations for the deformed factor and gauge coupling function. The second method is the DeWolfe-Gubser-Rosen (DGR) model~\cite{DeWolfe:2010he, DeWolfe:2011ts}, which numerically constructs a family of five-dimensional black holes. This model not only approximately matches equations of state and baryon susceptibilities with corresponding lattice QCD data~\cite{Karsch:2007dp} at zero chemical potential for $2+1$ flavor QCD matter but also reveals a line of first-order phase transitions terminating at a CEP located at $(\mu_{\text{CEP}}, T_{\text{CEP}})=(783, 143)$ MeV. Recent refinements to this model~\cite{Li:2020spf, Cai:2022omk} have enabled quantitative matching with up-to-date lattice data~\cite{HotQCD:2014kol, Borsanyi:2021sxv} at $\mu_B=0$ for $2+1$ flavor QCD matter, thereby determining the precise coordinates of the critical endpoint at $(\mu_{\text{CEP}}, T_{\text{CEP}})=(555, 105)$ MeV and characterizing the first-order transition line.  The location of CEP in $2+1$ flavor QCD has been confirmed in the model-independent approach \cite{Hippert:2023bel}. Further, the model parameters for pure SU(3) gauge theory have been determined in \cite{He:2022amv} through accurate matching with the latest lattice QCD data~\cite{Boyd:1996bx, Caselle:2018kap}, yielding a strong first-order confinement/deconfinement phase transition at $T_c = 276.5$ MeV, consistent with lattice QCD predictions. The phase diagram with rotation was examined in~\cite{Zhao:2022uxc}.

Experimentally, pinpointing the location of the CEP has been a keen focus. Yet, predicting it theoretically is challenging due to strong coupling properties in that region and the limitations of lattice techniques at finite chemical potential. Therefore, determining the CEP through a reasonable non-perturbative approach holds significant value. Furthermore, it is anticipated that the dynamic characteristics of the CEP, including critical exponents, align with the universality class of the 3D Ising model or the liquid/gas transition. Indeed, the critical exponents we derive for a $2$-flavor QCD system in the present study closely match those of the 3D Ising model and the liquid/gas transition, affirming this correspondence. Additionally, critical exponents have been calculated using other holographic models as well~\cite{Chen:2018vty, Chen:2018msc}, with outcomes that approach the predictions of the mean field theory. 

In this study, we employ holography to investigate the thermodynamic properties and dynamics of the CEP in $2$-flavor QCD matter. The Einstein-Maxwell-dilaton (EMD) gravity framework has been widely utilized in previous research to explore the QCD phase structure and other crucial physical quantities, as reviewed in recent works~\cite{Chen:2022goa, Rougemont:2023gfz}. By quantitatively aligning the behavior of relevant thermodynamic parameters with state-of-the-art lattice QCD data, we determine model parameters. This enables us to predict the CEP's location and delve into dynamic aspects by computing critical exponents near the CEP. Additionally, we utilize the self-consistent thermodynamic relations outlined in~\cite{Li:2020spf, Cai:2022omk} to analyze the variations in thermodynamic quantities, such as entropy density, pressure, trace anomaly, higher-order baryon number susceptibility, with increasing chemical potential.

The structure of this work is as follows: In Section~\ref{sec:01_model}, we establish a holographic QCD (hQCD) model featuring two flavors of light dynamical quarks, with all parameters determined based on state-of-the-art lattice QCD data at $\mu_B=0$~\cite{Burger:2014xga, Datta:2016ukp}. Section~\ref{sec:02_therm-quant} delves into a detailed analysis of thermodynamic quantities and certain order parameters at finite $\mu_B$, culminating in the construction of the $T$-$\mu_B$ phase diagram. We locate the CEP and compare its position with predictions from other low-energy effective models of QCD. In Section~\ref{sec:crit-exp}, we compute various critical exponents associated with the CEP and compare them with experimental results in non-QCD fluids, as well as with other models, including mean-field (van der Waals) theory, the full quantum 3D Ising model, and the DGR model~\cite{Goldenfeld:1992qy, DeWolfe:2010he}. We conclude with some discussion in Section~\ref{sec:summary}.

\section{Holographic QCD model}\label{sec:01_model}

We examine a five-dimensional bulk theory describing QCD using the Einstein-Maxwell-Dilaton (EMD) gravity framework. The action governing this system is specified in~\cite{Cai:2022omk}:
\begin{equation}\label{action}
  S_M =\frac{1}{2\kappa_N^2}\int d^5x\sqrt{-g}[\mathcal{R}-\frac{1}{2}\nabla_\mu\phi\nabla^\mu\phi-\frac{Z(\phi)}{4}F_{\mu\nu}F^{\mu\nu}-V(\phi)]\,,
\end{equation}
In this context, $\kappa_N^2$ stands for the effective Newton constant, while $g_{\mu\nu}$ represents the metric of the bulk spacetime. The field $\phi$ corresponds to the dilaton, responsible for breaking the conformal symmetry of the corresponding boundary theory. Additionally, $F_{\mu\nu}$ denotes the field strength tensor of the vector field $A_\mu$. This framework introduces two essential coupling functions, $Z(\phi)$ and $V(\phi)$. The former captures the equation of state (EOS) and sound velocity properties at zero chemical potential, while the latter is solely responsible for the behavior of baryon number susceptibilities (BNS) under the same conditions.

The hairy black holes take the following form~\cite{Li:2020spf, Cai:2022omk}:
\begin{align}\label{metric}
  ds^2 &=-f(r)e^{-\eta(r)}dt^2+\frac{dr^2}{f(r)}+r^2d\boldsymbol{x_3^2}\,,\nonumber\\
  \phi &=\phi(r),\quad\quad\quad A_\mu dx^\mu=A_t(r)dt\,,
\end{align}
with $d\boldsymbol{x_3^2}=d x_1^2+d x_2^2+d x_3^2$. The definition range of holographic radial coordinate $r$ is $[r_h, \infty)$, where the position of event horizon  $r_h$ is determined by $f(r_h)=0$ and the AdS boundary corresponds to $r\rightarrow \infty$. The Hawking temperature and entropy density are given by
\begin{equation}\label{temper}
  T=\frac{1}{4\pi}f'(r_h)e^{-\eta(r_h)/2},  \quad  s=\frac{2\pi}{\kappa_N^2}r_h^3\,.
\end{equation}
In order to obtain the configuration of hairy black holes, we need to numerically solve the equations of motion given by the variations of action~\eqref{action} under the ansatz~\eqref{metric} with appropriate boundary conditions (see~\cite{Li:2020spf, Cai:2022omk} for more technical details). Then the related thermodynamic quantities can be obtained by using the holographic renormalization.
\begin{figure}[!ht]
  \centering
  \includegraphics[width=0.45\textwidth]{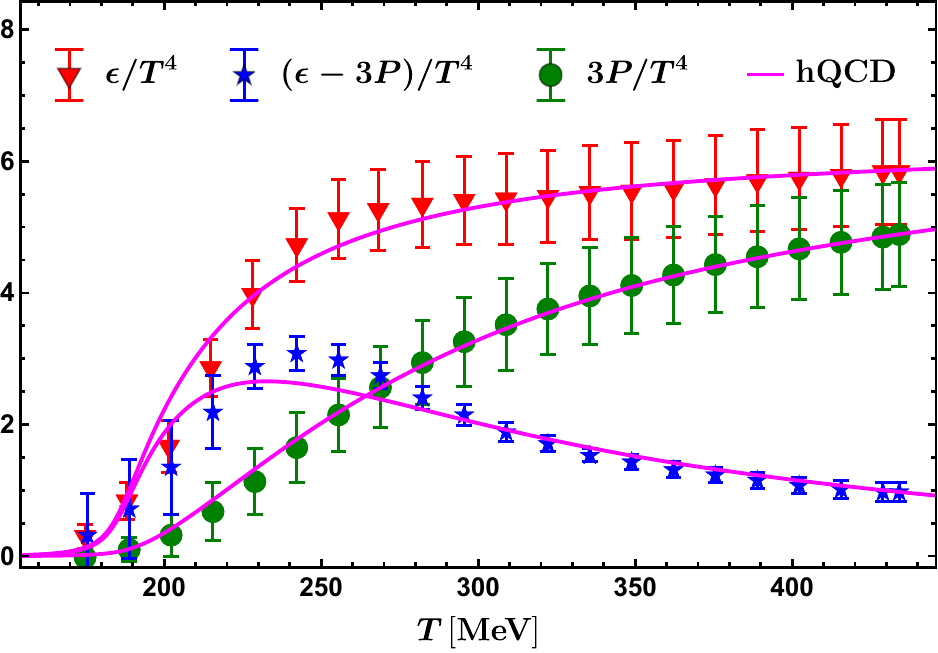}
   \includegraphics[width=0.5\textwidth]{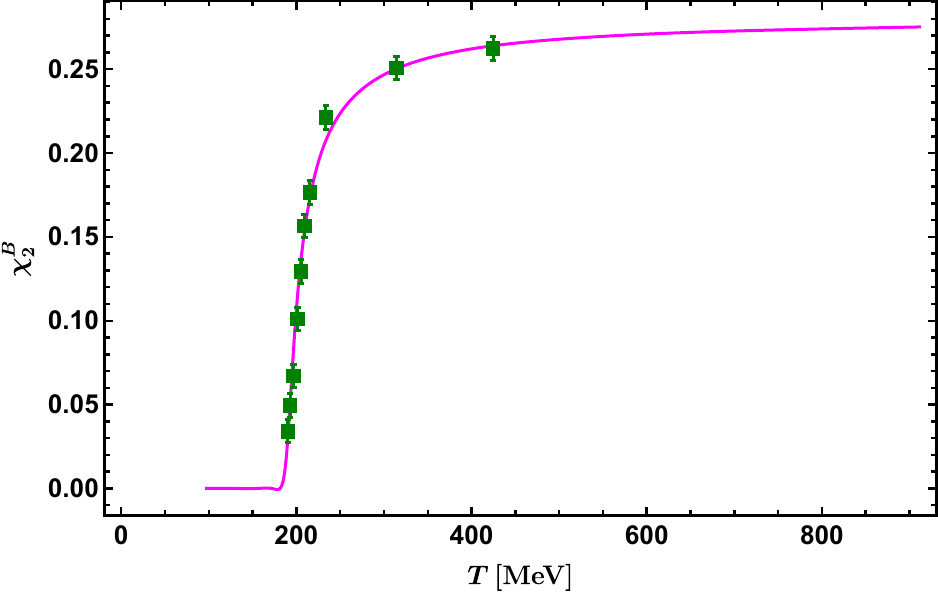}
  \caption{Comparison of thermodynamics at $\mu_B=0$ between our hQCD model (solid curves) and the lattice QCD (data with error bars).
  \textbf{Left panel}: The energy density $\epsilon$, pressure $P$ and trace anomaly (also called interaction measure) $\epsilon-3P$, as a function of temperature, where the lattice data comes from~\cite{Burger:2014xga}. \textbf{Right panel}: The temperature dependence of baryon number susceptibility $\chi^B_2$, where the lattice result is from~\cite{Datta:2016ukp}. }\label{latt-EOS}
\end{figure}
\begin{table}[!ht]
    \centering
    \setlength{\tabcolsep}{3.5mm}{
    \begin{tabular}{|c|c|c|c|c|c|c|c|c|}
     \cline{1-9}
    model  &  $c_1$  &  $c_2$  & $c_3$ & $c_4$  & $c_5$ & $\kappa_N^2$ & $\phi_s $(GeV) & b \\ \cline{1-9}
    pure $SU(3)$  & 0.735   &  0  &   &   &  & $2\pi(4.88)$ & 1.523 & -0.36458\\ \cline{1-9}
    2 flavor  & 0.710   &  0.0002  & 0.530  & 0.085  & 30 & $2\pi(3.72)$ & 1.227 & -0.25707\\ \cline{1-9}
    2+1 flavor  &  0.710  &  0.0037  & 1.935  & 0.085  & 30 & $2\pi(1.68)$  &  1.085 & -0.27341 \\ \cline{1-9} 
    \end{tabular}}
    \caption{Parameters for the pure $SU(3)$ gauge theory~\cite{He:2022amv}, $2$ flavor (this paper) and $2+1$ flavor models~\cite{Cai:2022omk} are obtained by matching the lattice QCD simulations. $\phi_s=r\phi|_{r\rightarrow\infty}$ is the source term that breaks the scale invariance of the dual system to essentially describe the real QCD dynamics. The parameter $b$ is from the holographic renormalization.}\label{model_para}
\end{table}

To better match the state-of-the-art lattice data, the potential and coupling function take the following structure~\cite{Cai:2022omk}, 
\begin{align}
     V(\phi) &=-12\cosh[c_1\phi]+(6c_1^2-\frac{3}{2})\phi^2+c_2\phi^6\,,\nonumber \\
     Z(\phi) &=\frac{1}{1+c_3}\text{sech}[c_4\phi^3]+\frac{c_3}{1+c_3}e^{-c_5\phi}\,,
\end{align}
with $c_1, c_2, c_3, c_4, c_5$ are free parameters. All parameters will be fixed by fitting the state-of-the-art lattice QCD data to well capture the behaviour of thermodynamic quantities for different physical systems. The values of these free parameters for different models are summarized in Table~\ref{model_para}, including the 2-flavor case in the present study.\,\footnote{The value of $c_4$ for $(2+1)$-flavor model has been made a slight modification from $0.085$ to $0.091$ to match the higher-order baryon number susceptibilities~\cite{Li:2023mpv}} One can find $Z(\phi)=0$ for the pure SU(3) model and the parameters $(c_1, c_4, c_5)$, \emph{i.e.} the coefficients of odd power of dialton $\phi$, keep unchanged for the finite quark flavor models. What we need to emphasize is that our current model mainly focuses on the dynamical properties of quarks and the quark flavor dynamics are effectively adopted into our five free parameters in $V$ and $Z$. All these parameters will be fixed by matching with cutting-edge lattice QCD data, for which the quantum characteristics of u, d quarks (such as isospin, spin, etc.) also are captured by these parameters.

\begin{figure}[!t]
  \centering
  \includegraphics[width=0.49\textwidth]{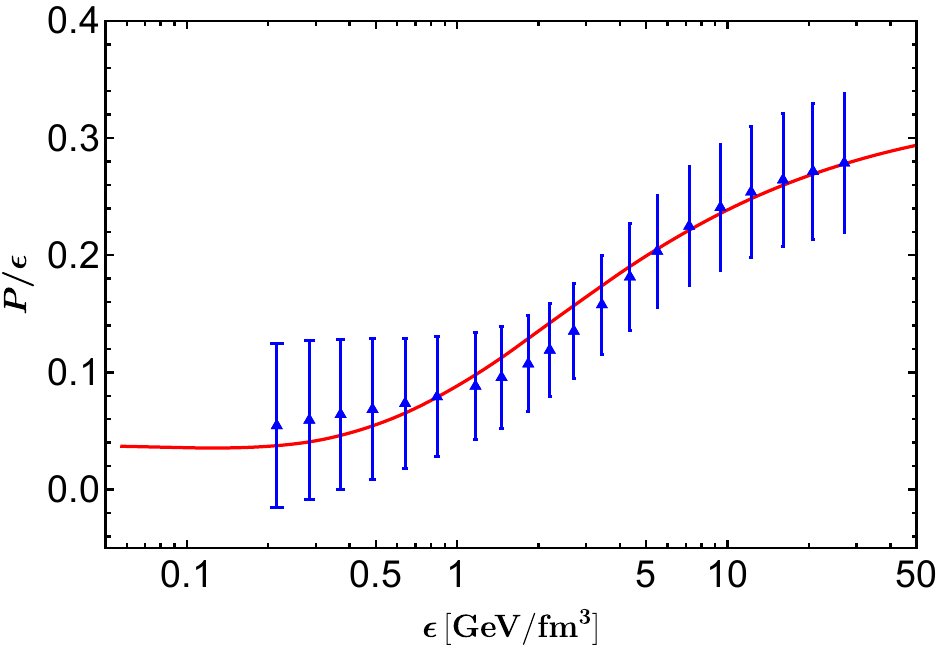}
  \includegraphics[width=0.49\textwidth]{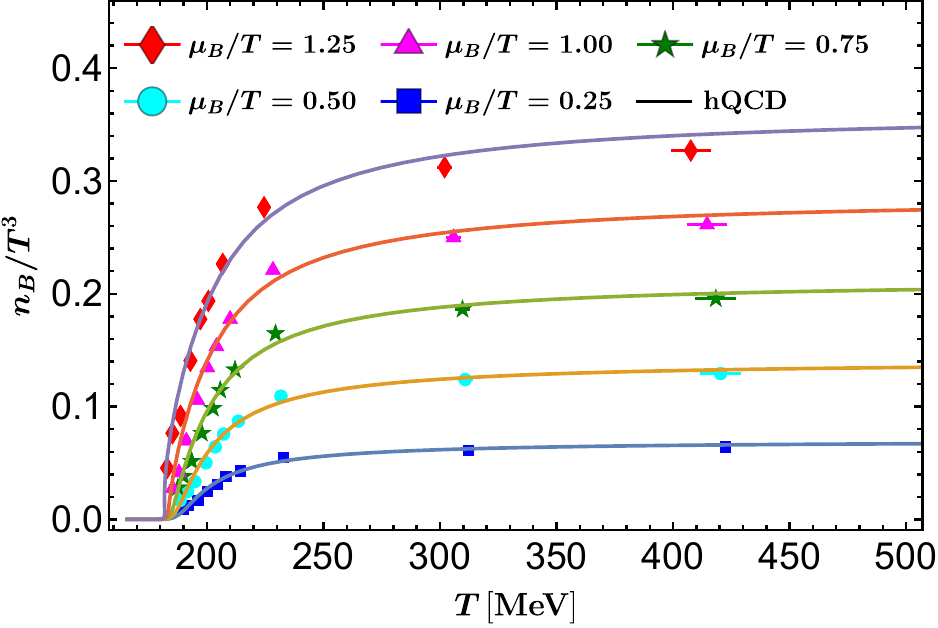}
  \caption{\textbf{Left panel}: The ratio of pressure and energy density $P/\epsilon$ versus energy density $\epsilon$ at $\mu_B=0$ with the lattice data being from~\cite{Burger:2014xga}. \textbf{Right panel}: The baryon number density $n_B$ as a function of temperature $T$ at fixed $\mu_B/T$, where the lattice data is from~\cite{Datta:2016ukp}. Our holographic results are all denoted by solid lines.}\label{latt-rho}
\end{figure}

For the 2-flavor model, we compare different thermodynamic quantities from our holographic setup with lattice simulation\,\footnote{The lattice data~\cite{Burger:2014xga} we used is from the simulations that have been carried out at the bare quark masses corresponding to pion masses $m_\pi \sim 360$ MeV and $N_t=12$ with $N_f=2$ degenerate quark flavor. In addition, to match the lattice simulation, we take the pseudo-critical temperature of the lattice simulation~\cite{Datta:2016ukp} as $T_c(\mu_B=0)=205$ MeV, which is within the deconfinement range of $219\pm3\pm14$ obtained by~\cite{Burger:2014xga}. It should be noted that the simulations from~\cite{Datta:2016ukp} were carried out using two flavors of dynamical staggered quarks with $m_\pi/m_\rho \sim 0.4$ and $N_t=8$.\label{f1:myfootnote}} at $\mu_B=0$ in Fig.~\ref{latt-EOS}. One can find that the temperature dependence of those quantities agrees well with lattice results, where the baryon number susceptibility $\chi^B_2$\,\footnote{Note that $\chi^B_2$ denoting dimensionless quantity in this paper is equal to $\chi_B^2/T^2$ of~\cite{Datta:2016ukp}.} at vanishing chemical potential is defined as $\chi^B_2(\mu_B=0)=\lim\limits_{\mu_B \to 0}\frac{1}{T^2}\frac{n_B}{\mu_B}$ with $n_B$ the baryon number density. In addition, as holographic predictions, we calculate the ratio of pressure and energy density as a function of energy density at zero chemical potential and the baryon number densities versus temperature for different $\mu_B/T$ ratios in Fig.~\ref{latt-rho}. The results show that the holographic predictions are in quantitative agreement with the lattice results\,\footnote{ Here we take the pseudo-critical temperature of deconfinement as $T_c=(204,\,203,\,202,\,200,\,196)$ MeV, corresponding to $\mu_B/T=(0.25,\,0.5,\,0.75,\,1.00,\,1.25)$, respectively. } available for small chemical potentials, which strongly supports our hQCD model, and thus, in terms of the thermodynamic properties of QCD, our holographic model surpasses the mean field level.

\section{Thermodynamics quantities and phase diagram}
\label{sec:02_therm-quant}

Having established the $N_f=2$ holographic model, we investigate thermodynamic properties and construct the phase diagram at finite $\mu_B$. It's important to note that all relevant thermodynamic quantities have been rigorously defined through holographic renormalization, as extensively detailed in~\cite{Li:2020spf, Cai:2022omk, Zhao:2022uxc}, and are not presented here for brevity.
\begin{figure}
  \centering
  \includegraphics[width=0.49\textwidth]{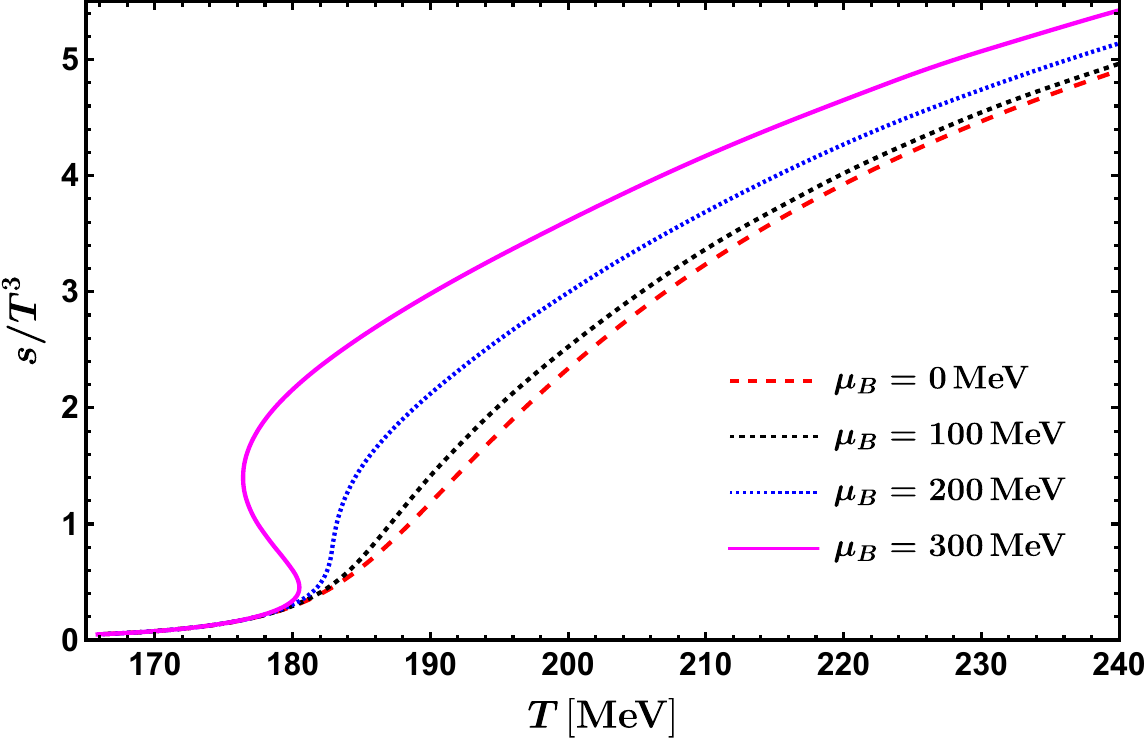}
  \includegraphics[width=0.49\textwidth]{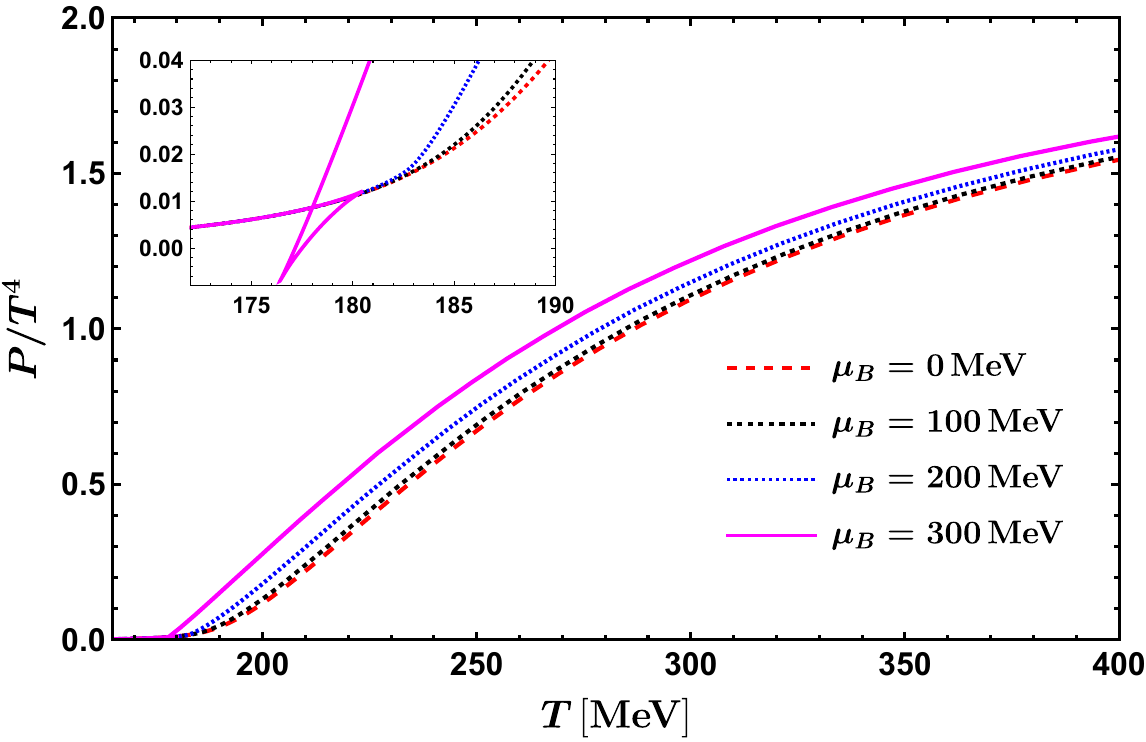}\\
  \includegraphics[width=0.49\textwidth]{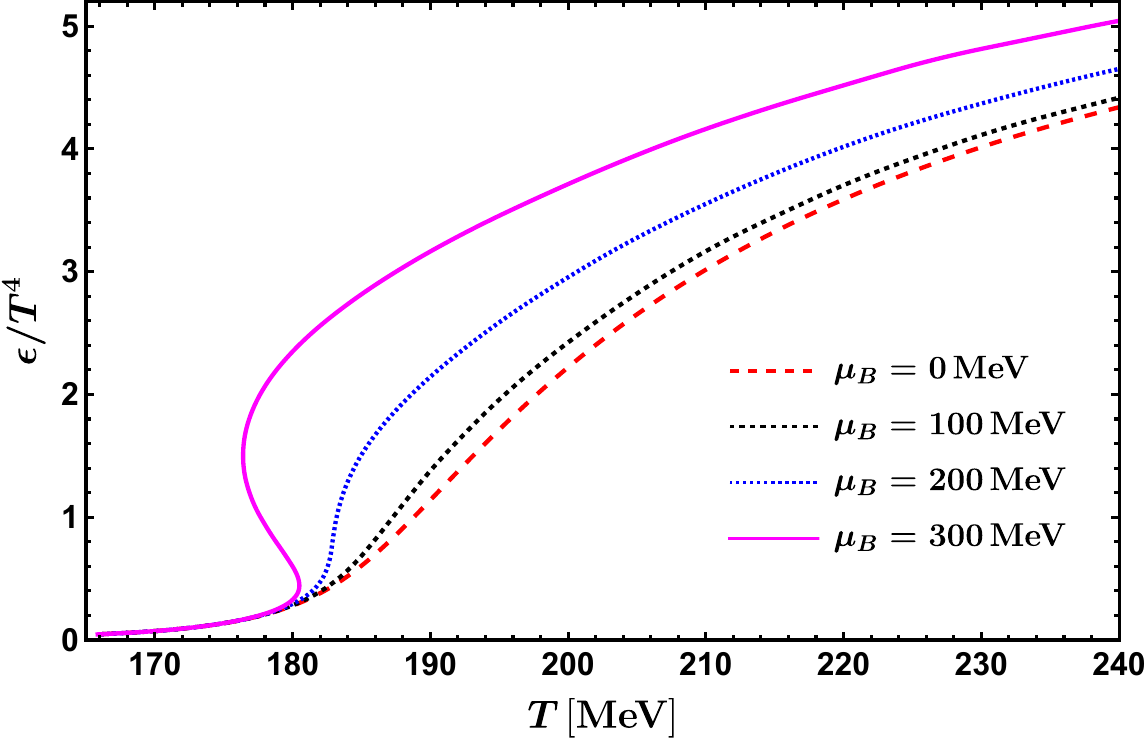}
  \includegraphics[width=0.49\textwidth]{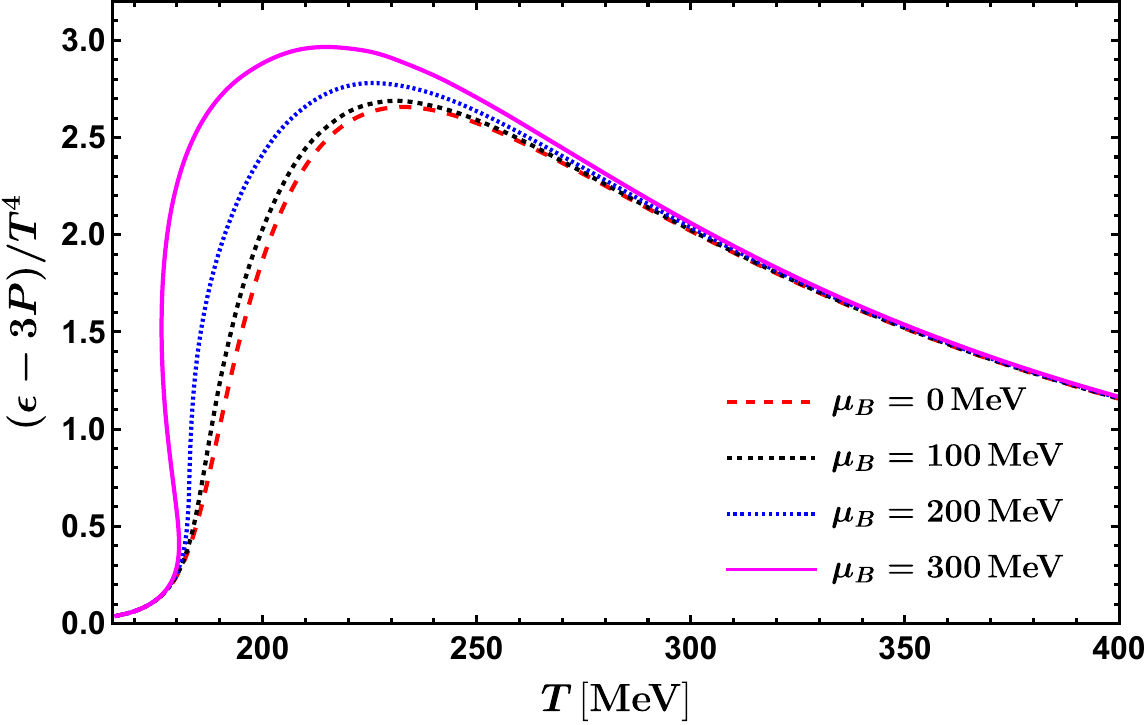}\\
  \caption{The entropy density $s$, pressure $P$, energy density $\epsilon$ and trace anomaly $\epsilon-3P$ as a function of $T$ at different values of $\mu_B$. These quantities are all enhanced by increasing the chemical potential.}\label{speI}
\end{figure}

In Fig.~\ref{speI}, we illustrate the temperature dependence of the Equation of State (EOS) and trace anomaly across various chemical potentials. 
As the increase of chemical potential, these quantities change from a single-valued behavior to a multi-valued one, marking the beginning of a first-order phase transition and the end of the crossover. The critical temperature of the first-order transition can be determined from the pressure $P$, which is nothing but the minus of the free energy density of our system. More precisely, the thermodynamically favored phase has the lowest free energy density. Thus, the critical temperature corresponds to the tip of the swallowtail in the temperature dependence of $P$, see the subset of the second plot of Fig.~\ref{speI}.
\begin{figure}
  \centering
  \includegraphics[width=0.49\textwidth]{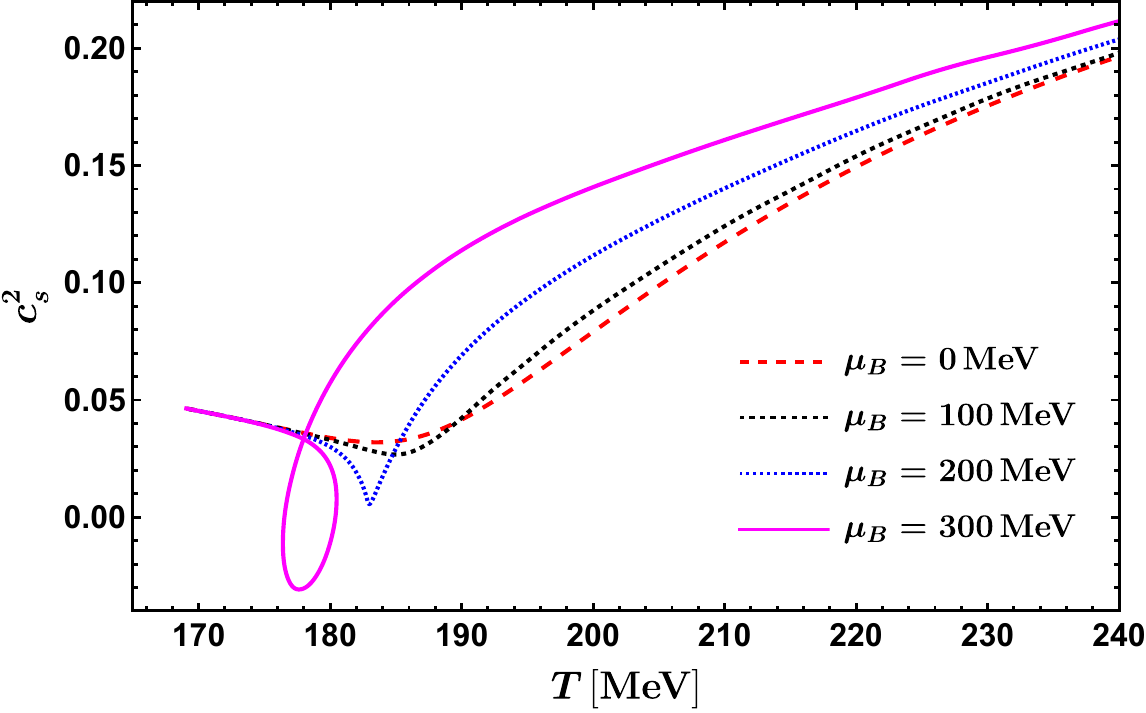}
  \includegraphics[width=0.49\textwidth]{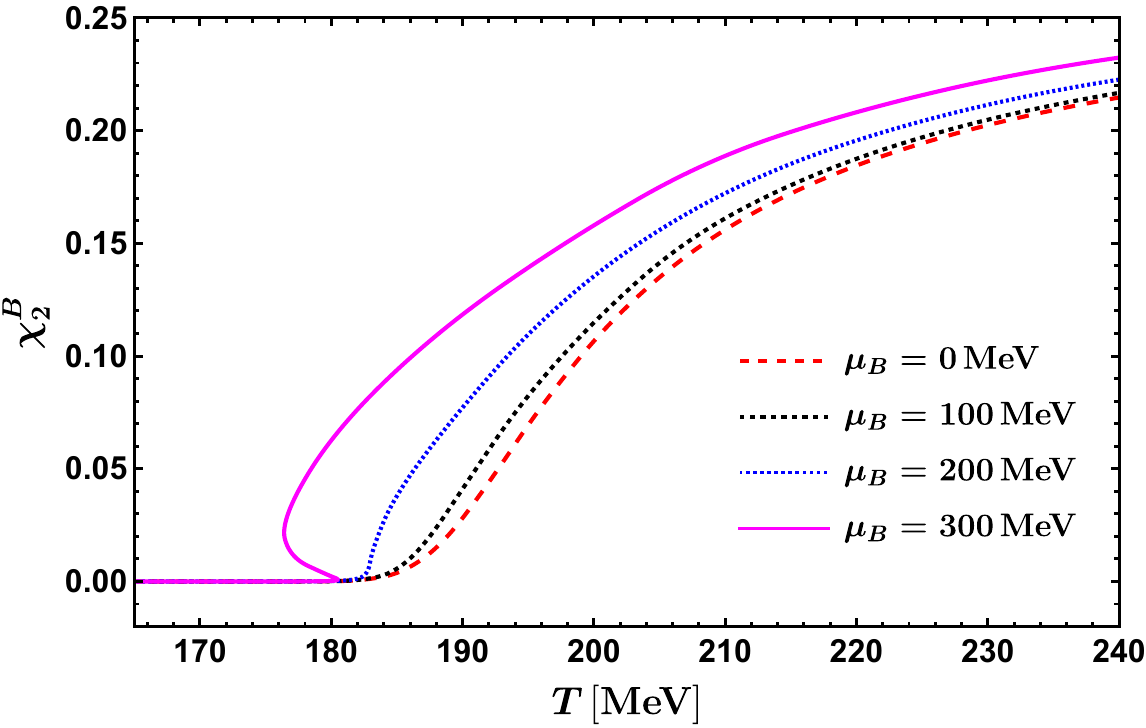}
  \caption{The squared speed of sound $c_s^2$ and baryon number susceptibility $\chi^B_2$ as a function of temperature at different chemical potentials. At small $\mu_B$, there is only a crossover. For sufficiently large $\mu_B$, a first-order phase transition is triggered.}\label{cschi}
\end{figure}

In the crossover region between the hadron resonance gas and the quark-gluon plasma (QGP), there is no unique way to determine the transition temperature in the literature. Nevertheless, one can define a pseudo-transition temperature to construct a comprehensive QCD phase diagram. This can be accomplished by identifying key indicators such as the minimum squared speed of sound, the inflection point of the second-order baryon number susceptibility, or the susceptibility of the Polyakov loop. These indicators capture the pronounced change in degrees of freedom between the QGP and the hadron resonance gas. In Fig.~\ref{cschi}, we present the behavior of the squared speed of sound $c_s^2(T,\mu_B)=\partial P/\partial \epsilon$ (left panel) and the baryon number susceptibility $\chi_2^B(T,\mu_B)=(\partial n_B/\partial {\mu_B})/T^2$ (right panel) for different $\mu_B$. At low chemical potentials, the single-valued behavior indicates a smooth crossover. Notably, both $c_s^2$ and $\chi_2^B$ exhibit enhancement as the chemical potential increases.
\begin{figure}
  \centering
  \includegraphics[width=0.49\textwidth]{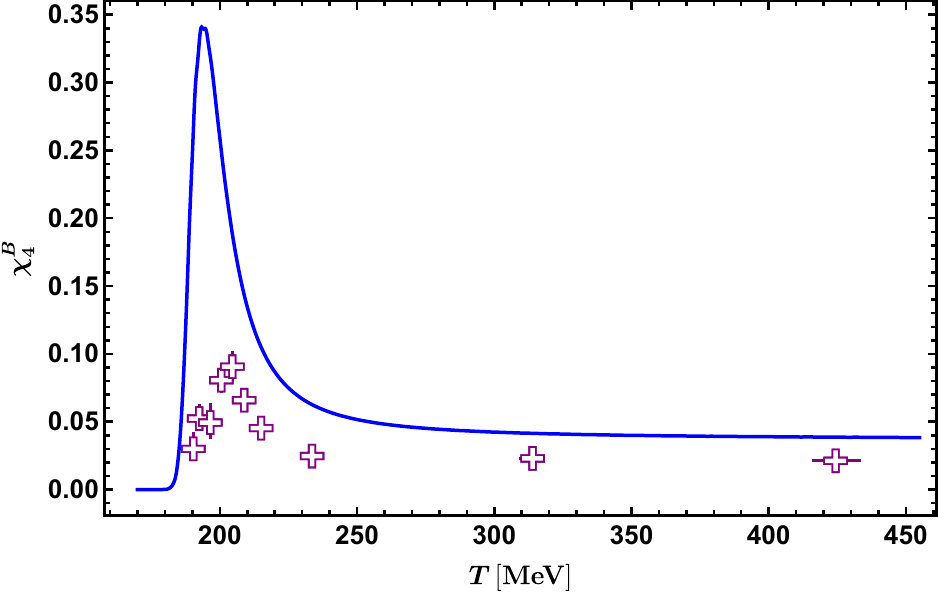}
  \includegraphics[width=0.49\textwidth]{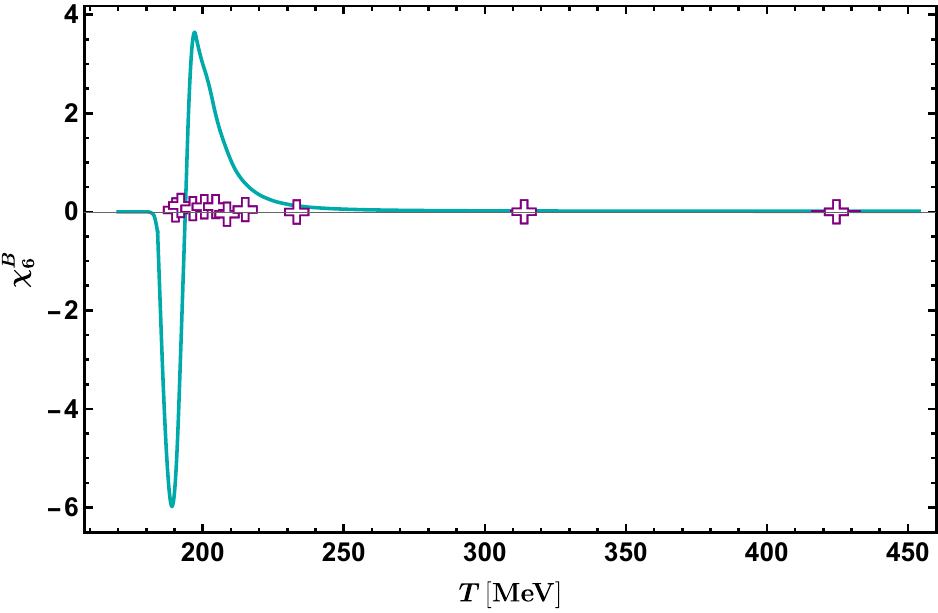}
  \caption{Higher order baryon number susceptibilities $\chi^B_4$ (left) and $\chi^B_6$ (right) as a function of temperature at $\mu_B=0$. The holographic results of susceptibilities are qualitatively consistent with the lattice data~\cite{Datta:2016ukp}. }\label{chin}
\end{figure}

In addition, it is also of great significance to study the higher-order baryon number susceptibilities defined as the $n$-th order derivatives of the pressure concerning the baryon chemical potential.
\begin{equation}\label{BNS}
   \chi^B_n=\frac{\partial^n(P/T^4)}{\partial(\mu_B/T)^n}.
\end{equation}
The $\mu_B$ dependence of pressure excess $\Delta P(\mu_B,T)=P(\mu_B,T)-P(0,T)$ can easily be represented by a Taylor series~\cite{Bazavov:2017dus}
\begin{equation}
    \Delta P(\mu_B, T)/T^4=\sum_{n=1}^{\infty} \frac{\chi_{2n}^B|_{\mu_B=0}}{(2n)!}\left( \frac{\mu_B}{T}\right)^{2n}.
\end{equation}
Note that the odd-order baryon number susceptibilities vanish at $\mu_B=0$, \emph{i.e.} $\chi_{2k+1}^B(\mu_B=0,T)=0$ due to the CP symmetry. Moreover, the ratios of baryon number fluctuations~\cite{Li:2023mpv, Isserstedt:2019pgx} 
emerge as a potent tool to probe the phase transitions. These ratios correspond to the corresponding ratios of cumulants derived from experimental data accessible through event-by-event analyses of heavy-ion collisions. For example, 
\begin{equation}
 \frac{\chi^B_4}{\chi^B_2}=\kappa_B\sigma_B^2, \quad  \frac{\chi^B_3}{\chi^B_2}=S_B\sigma_B, \quad \frac{\chi^B_1}{\chi^B_2}=\frac{M_B}{\sigma_B^2},  
\end{equation}
where $\kappa_B$, $\sigma_B^2$, $S_B$, and $M_B$ denote the kurtosis, variance, skewness, and mean of the net-baryon distribution, respectively (see~\cite{Luo:2017faz, Bzdak:2019pkr, Asakawa:2015ybt} for more details). 

In Fig.~\ref{chin}, we present the numerical results for the higher-order baryon number susceptibilities at $\mu_B = 0$. We also compare these susceptibilities ($\chi^B_4$ and $\chi^B_6$) and the outcomes from state-of-the-art lattice QCD simulations. Obviously, near the pseudo-critical temperature, the values of these magnetic susceptibilities will increase rapidly. It is worth noting that the holographic results for $\chi^B_4$ and $\chi^B_6$ show qualitative consistency with the lattice data, and any quantitative differences may be attributed to the factors detailed in footnote\textsuperscript{\ref{f1:myfootnote}}.

\begin{figure}
  \centering
  \includegraphics[width=0.49\textwidth]{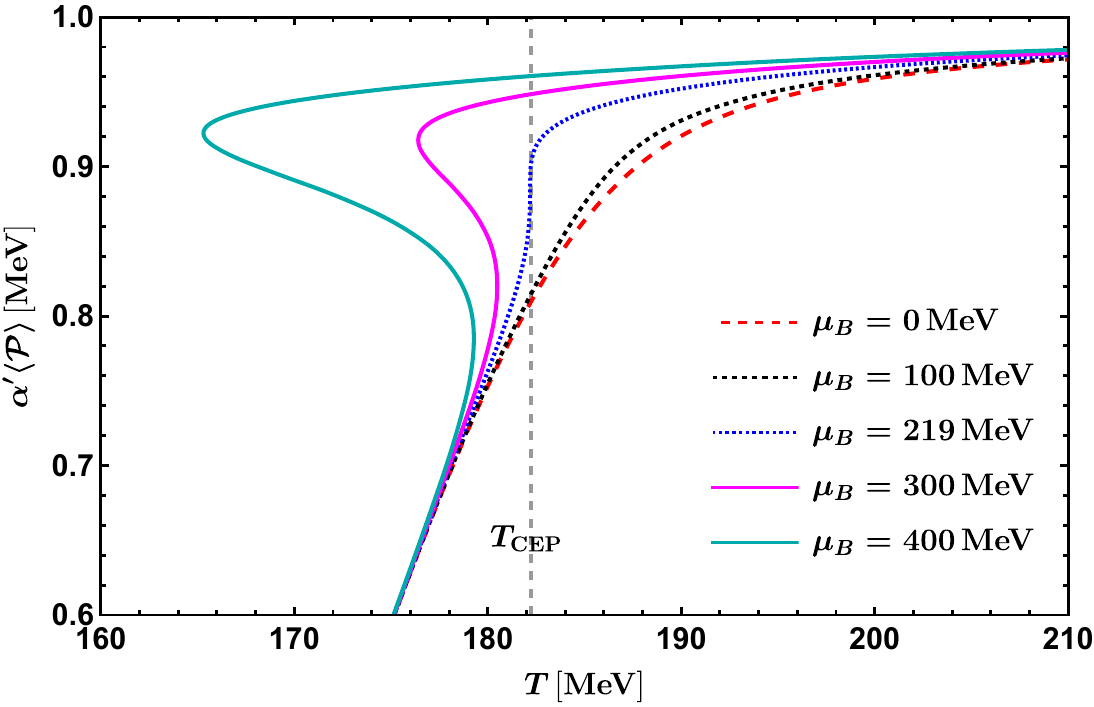}
  \includegraphics[width=0.49\textwidth]{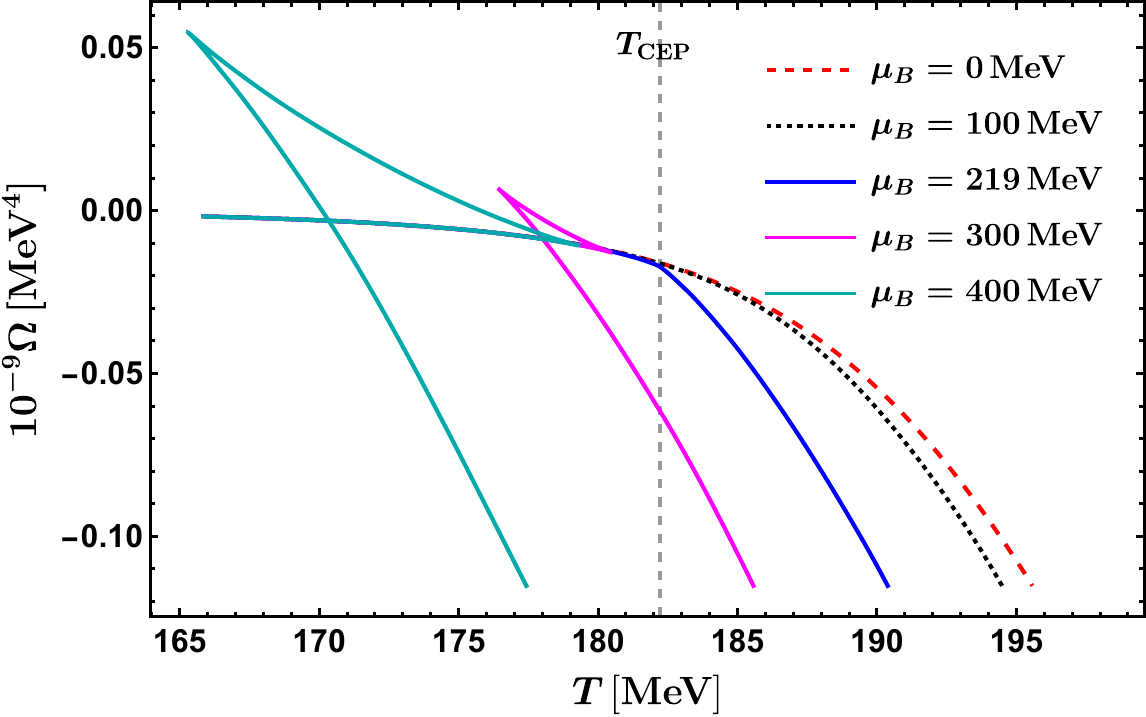}
  \caption{The Polyakov loop $\left<\mathcal{P}\right>$ (left) and free energy density $\Omega$  at different $\mu_B$. The phase transition becomes first-order when $\mu_B>219\,\text{MeV}$.
 }\label{polyafree}
\end{figure}

Continuing with the previous content, we further examine the temperature dependence of the Polyakov loop $\left<\mathcal{P}\right>$ in the left panel of Fig.~\ref{polyafree}. While the Polyakov loop is not an ideal order parameter for the 2-flavor QCD due to the influence of quark degrees of freedom that disrupt the $Z(N_c)$ symmetry, it could be an effective order parameter in this case. One finds that $\left<\mathcal{P}\right>$ exhibits a non-zero value in the low-temperature phase, followed by a rapid increase as the temperature approaches the pseudo-transition region. As $\mu_B$ approaches the critical value $\mu_B=219\,\text{MeV}$ from below, the susceptibility of $\left<\mathcal{P}\right>$ becomes infinite. Moreover, for $\mu_B>219\,\text{MeV}$, $\left<\mathcal{P}\right>$ develops a multi-valued behavior, suggesting a first-order phase transition. The corresponding behavior of the free energy $\Omega$ versus temperature is presented in the right panel of Fig.~\ref{polyafree}. The temperature dependence of $\Omega$ decreases smoothly for $\mu_B<219\,\text{MeV}$, while it becomes a swallowtail for $\mu_B>219\,\text{MeV}$, signaling a first-order phase transition. The location of the CEP where the swallowtail terminates is found to be at $(\mu_{\text{CEP}},\, T_{\text{CEP}})=(219\,\text{MeV}, \, 182\, \text{MeV})$, which is consistent with the result from the Polyakov loop analysis.
\begin{figure}
  \centering
  \includegraphics[width=0.9\textwidth]{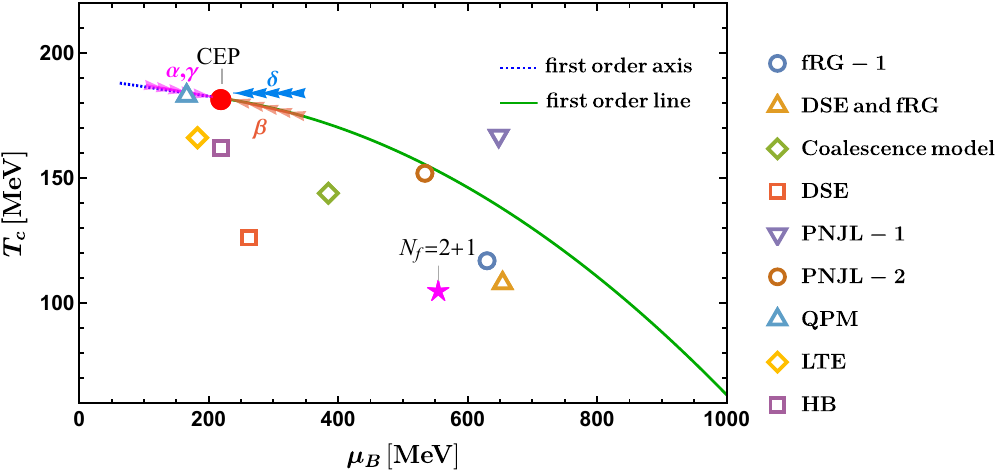}\\
  \caption{The phase diagram of QCD matter in our 2-flavor holographic model. The green curve shows the phase boundary for the first-order phase transition, and the blue dash line denotes the first-order axis. The green first-order line terminates at the CEP $(\mu_{\text{CEP}},\, T_{\text{CEP}})=(219\,\text{MeV}, \, 182\, \text{MeV})$ (red point). The location of CEP predicted by other approaches are presented as well, including functional renormalization group (fRG), Schwinger–Dyson equations (DSE), the combination of functional renormalization group (fRG) and Schwinger–Dyson equations (DSE), Nambu-Jona-Lassinio effective chiral model coupled to the Polyakov loop (PNJL), quasiparticle model (QPM), lattice taylor expansion (LTE), hadronic bootstrap (HB), and the coalescence model for light nuclei production. fRG-1 is from~\cite{Fu:2019hdw}. DSE and fRG is from~\cite{Gao:2020qsj}. Coalescence model is from~\cite{Sun:2018jhg}. DSE is from~\cite{Gao:2016qkh}. PNJL-1 is from~\cite{McLerran:2008ua} and PNJL-2 is from~\cite{Sasaki:2010jz}. QPM is from~\cite{Srivastava:2010xa}. LTE is from~\cite{Gavai:2004sd}. HB is from~\cite{Antoniou:2005gn}. The magenta star represents the CEP of $2+1$ flavor QCD obtained by our previous model in~\cite{Cai:2022omk}. We also indicate the directions of approach of the various critical exponents.
}\label{PTD}
\end{figure}

Having comprehensively examined all thermodynamic quantities, we construct the phase diagram for 2-flavor QCD matter regarding temperature and baryon chemical potential, as depicted in Fig.~\ref{PTD}. The green curve denotes the phase boundary for the first-order phase transition, uniquely determined by the characteristic swallowtail behavior of the free energy. The blue dashed line represents the tangent of the first-order phase transition line at the CEP, which will be called the first-order axis. The location of the CEP $(\mu_{\text{CEP}},\, T_{\text{CEP}})=(219\,\text{MeV}, \, 182\, \text{MeV})$ is marked with the red point. Therefore, the smooth crossover between the hadronic phase of color-neutral bound states at low $T$ and small $\mu_B$, and the QGP at high $T$ and large $\mu_B$ transforms a first-order transition with increasing chemical potential. Moreover, the critical temperature decreases as $\mu_B$ is increased. An interesting point is that the location of the CEP from our holographic theory is at $\mu_\text{CEP}/T_\text{CEP}=1.2$, while the extrapolation of lattice data yields the CEP in $1.5\pm 0.2 \leq \mu_\text{CEP}/T_\text{CEP} \leq 1.85\pm0.04$~\cite{Gupta:2014qka, Datta:2016ukp}. From the right panel of Fig.~\ref{latt-rho}, it can be observed that the baryon number densities from holography align very well with those from lattice calculations when $\mu_B/T<1$. However, above this threshold, it appears that lattice techniques do not work as effectively, leading to a slight discrepancy with the holographic results. We also include the location of CEP predicted by other approaches. Our CEP is relatively close to those predicted by QPM~\cite{Srivastava:2010xa}, LTE~\cite{Gavai:2004sd}, and HB~\cite{Antoniou:2005gn}. Specifically, the CEP predicted by LTE is at  $\mu_\text{CEP}/T_\text{CEP}\approx1.1$, which is lower than our predicted value, $\mu_\text{CEP}/T_\text{CEP}=1.2$. Therefore, the lattice group could easily validate our prediction for CEP by either employing different methods that work well up to $\mu_B/T<1.25$ or by accumulating a large amount of data around $\mu_B/T=1.2$. Furthermore, we show the location of the CEP at $(\mu_{\text{CEP}},\, T_{\text{CEP}})=(555\,\text{MeV}, \, 105\, \text{MeV})$ predicted by our 2+1 flavor holographic model~\cite{Cai:2022omk}. Notably, the substantial influence of dynamical quark flavors on the location becomes apparent. The phase diagram of 2-flavor holographic QCD was also qualitatively studied in~\cite{Chen:2020ath, Chen:2019rez} using the potential reconstruction method. There is no first-order deconfinement phase transition in the $T-\mu_B$ plane, while there develops a first-order chiral phase transition as $\mu_B$ is increased.

\section{Critical phenomena near the CEP}\label{sec:crit-exp}
\begin{figure}
  \centering
  \includegraphics[width=0.32\textwidth]{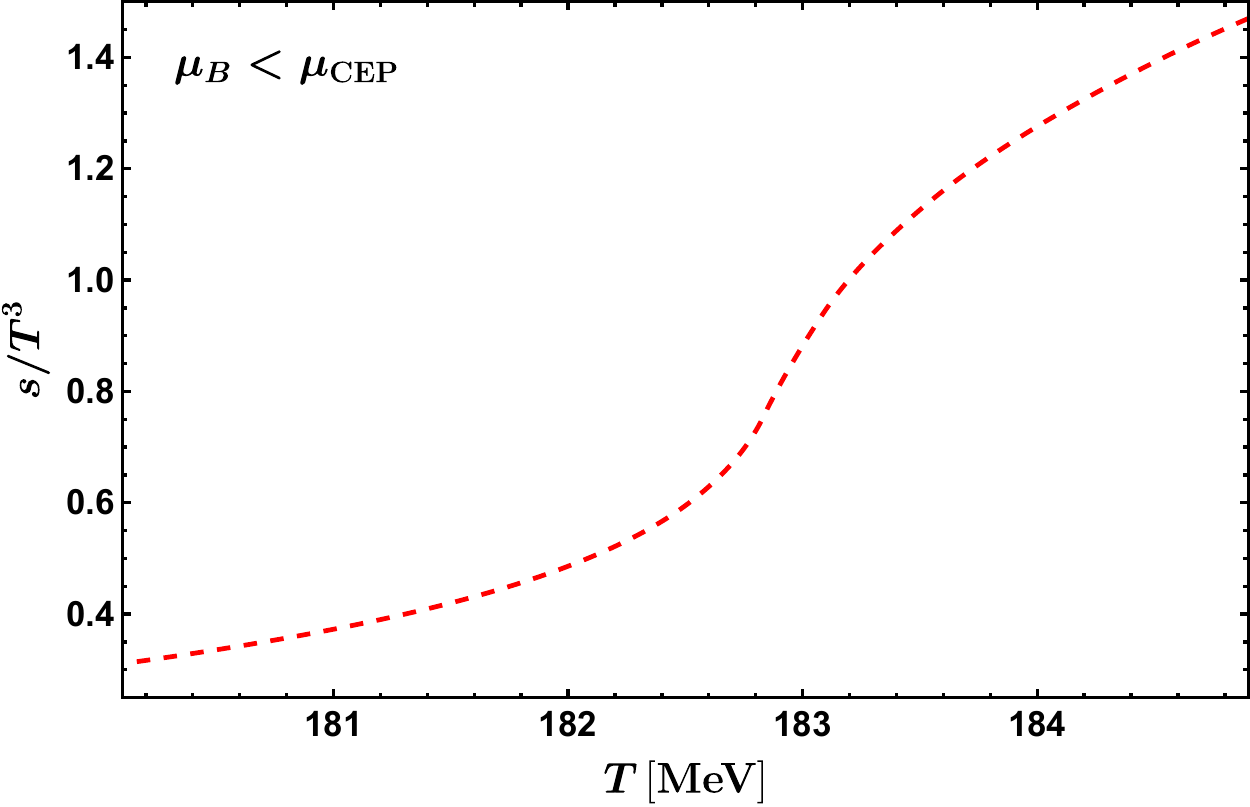}
  \includegraphics[width=0.32\textwidth]{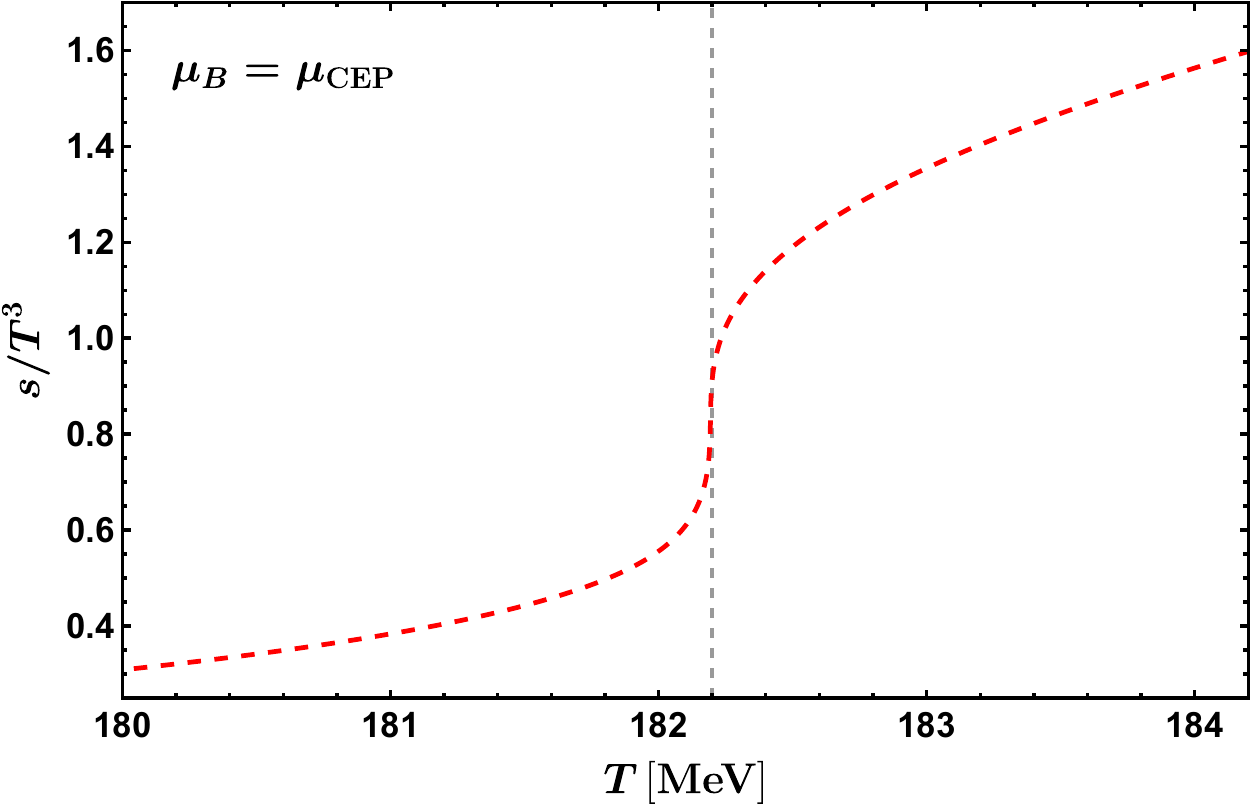}
  \includegraphics[width=0.32\textwidth]{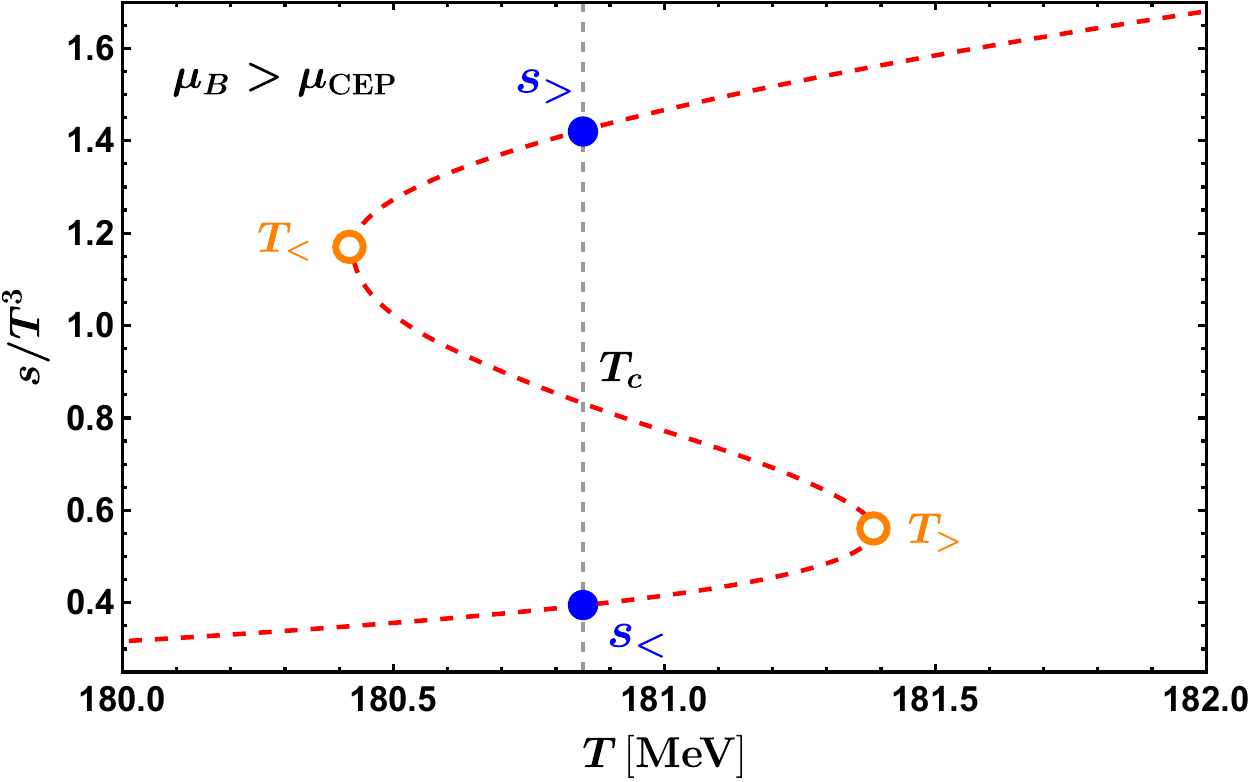}\\
  \caption{The entropy density $s$ as a function of $T$ for several values of $\mu_B$ near the CEP. For $\mu_B<\mu_{\text{CEP}}$, the curve $s(T)$ is single-valued (left), while for $\mu_B>\mu_{\text{CEP}}$ it becomes multi-valued (right). At $\mu_B=\mu_{\text{CEP}}\, \text{and} \;T=T_{\text{CEP}}$, the slope is infinite (middle).}\label{s3}
\end{figure}

Near the vicinity of CEP, the behavior of thermodynamic quantities usually follows the power laws characterized by critical exponents.
These exponents are universal, meaning they show the same values in different physical systems undergoing phase transitions, regardless of the details of the system. They are at the heart of the study of critical phenomena.
Among the six widely recognized thermodynamic critical exponents, $\alpha, \beta, \gamma, \delta, \nu, \eta$, the present study focuses on $\alpha, \beta, \gamma, \delta$ which will be discussed in detail below. The remaining two, $\nu$ and $\eta$, require spatial correlation functions and are not discussed here.

To determine the value of a critical exponent, it is necessary to determine the axis of interest near the CEP. This axis is commonly defined as the first-order line, the first-order axis, or the critical isotherm. Because the thermodynamic quantities near the CEP follow some power law behavior, a log-log plot is employed for analysis. The critical exponents can be deduced from the slope of the straight-line approximation. In practice, linear regression via least squares will be used to determine these slopes consistently throughout this section.

To calculate the critical exponents, a thorough examination of thermodynamic quantities in different transition regions is necessary. Entropy density serves as an example here. In Fig.~\ref{s3}, we show the behavior of entropy density with temperature for three cases: $\mu_B < \mu_{\text{CEP}}$ (left panel), $\mu_B = \mu_{\text{CEP}}$ (middle panel), and $\mu_B > \mu_{\text{CEP}}$ (right panel). For the first case with a constant chemical potential $\mu_B < \mu_{\text{CEP}}$, the isopotential line avoids the first-order line decipted by the green curve of Fig.~\ref{PTD}, yielding a unique value of entropy density $s$ at each temperature. In contrast, when $\mu_B > \mu_{\text{CEP}}$, the isopotential intersects the first-order line, resulting on a multi-valued entropy density around $T_{\text{CEP}}$. This behavior resembles an ``S"-curve as $T$ increases, characterized by the existence of three branches of states at the same point in the phase diagram. As visible from the right panel of Fig.~\ref{s3}, there are two inflection points $T_<$ and $T_>$, \emph{i.e.} the locations of the local minimum and maximum of the isopotential curve $s(T)$. The critical temperature $T_c \approx (T_< + T_>)/2$ as the critical point is approached. It is manifest that the middle branch lying in between the upper and lower branches has a negative specific heat $C_{v}=T(\partial s/\partial T)|_{\mu_B}$ and thus corresponding to thermodynamically unstable states. For later convenience, we denote $s_>$ and $s_<$ as the value of entropy density at $T_c$ for the upper and lower branches, respectively. When $\mu_B = \mu_{\text{CEP}}$, these three branches merge into one, casing the infinite slope of the curve $s(T)$ on the critical isopotential (see the middle panel of Fig.~\ref{s3}). This suggests the divergence of specific heat $C_v$ at the CEP. In practice, we obtain the entropy density at the CEP $s_{\text{CEP}}$ as the converging point of both $s_>$ and $s_<$ as they approach the CEP, which will be discussed further in subsection \ref{sec:beta}.

\begin{figure}[!t]
  \centering
  \includegraphics[width=0.65\textwidth]{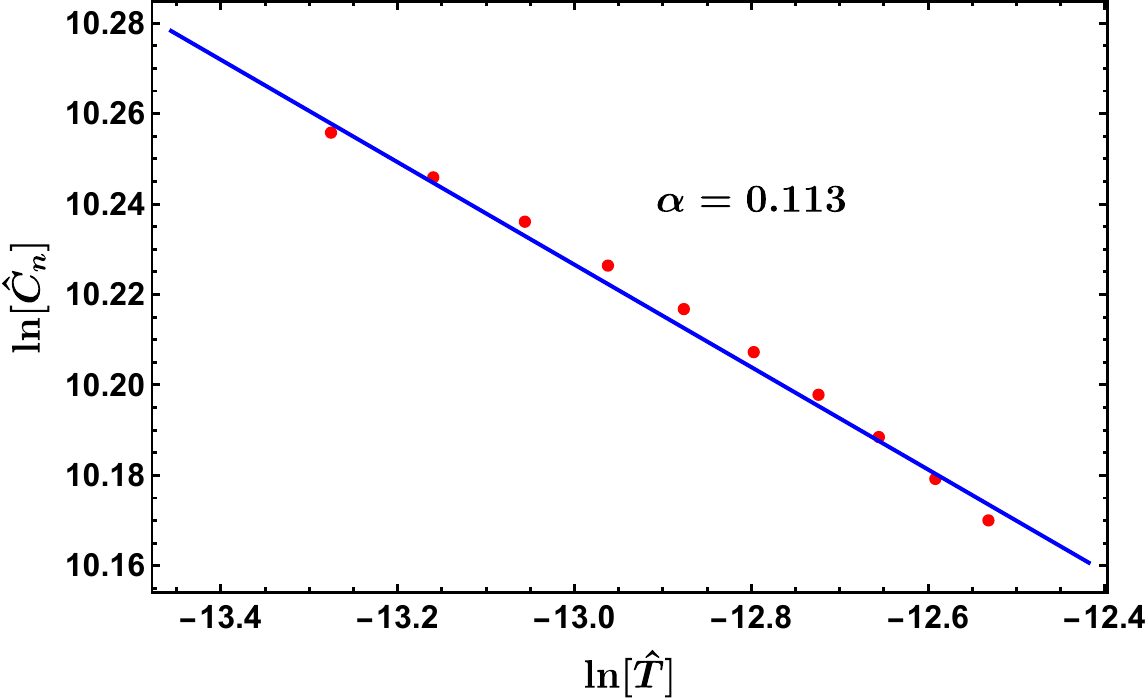}
  \caption{The dimensionless specific heat $\hat{C}_{n}=C_{n}/T^3$ in a log-log plot with $\hat{T}=\frac{T-T_{\text{CEP}}}{T_{\text{CEP}}}$ near the critical endpoint along the first-order axis. The slope of the best-fit line to our data (blue line) yields $\alpha=0.113$.}\label{fig-alpha}
\end{figure}
\begin{figure}
  \centering
  \includegraphics[width=0.65\textwidth]{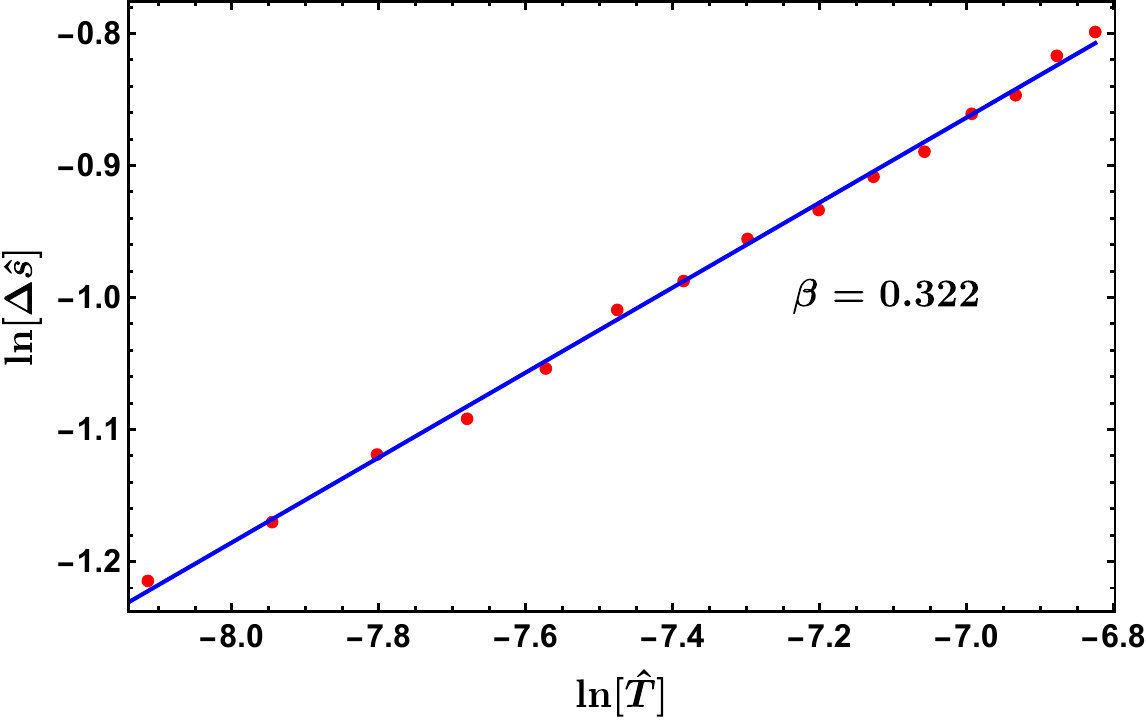}
  \caption{The discontinuity in the dimensionless entropy density $\hat{s}=s/T^3$ as one approaches the CEP on a log-log plot. The value of $\beta$ obtained from the slope is $\beta$= 0.322. }\label{fig-beta}
\end{figure}
%

\subsection{Critical exponent-\texorpdfstring{$\alpha$}{alpha} along first order axis}\label{sec:alpha}

The first-order line ends at the CEP. 
Near the critical {end}point {along the axis defined by the first-order axis}, the exponent $\alpha$ characterizes the power law pattern of the specific heat at constant $n_B$,
\begin{equation}\label{alpha}
    C_{n}\equiv T\left(\frac{\partial s}{\partial T}\right)_{n_B}=-T\left( \frac{\partial^2\Omega}{\partial T^2}-\frac{(\partial^2\Omega/\partial T\partial\mu)^2}{(\partial^2 \Omega/\partial \mu^2)}\right)\sim|T-T_{\text{CEP}}|^{-\alpha}.
\end{equation}
To sidestep the intricacies of the first-order line, we opt to approach the CEP from the crossover region where $\mu_B<\mu_{\text{CEP}}$.
A benefit in computation is that the constant $n_B$ line nearly aligns with the first-order axis, which has been used in holography to calculate the critical exponents $\alpha$ and $\gamma$, see \emph{e.g.}~\cite{DeWolfe:2010he}. 

In Fig.~\ref{fig-alpha}, we show the temperature dependence of $C_{n}$ near the CEP along the first-order axis. The power law~\eqref{alpha} is manifest in the log-log plot. It shows a weak divergence with
\begin{equation}
    \alpha=0.113.
\end{equation}
Our holographic result is very close to that of the experiments in non-QCD fluids and the the full quantum 3D Ising model quantitatively~\cite{Goldenfeld:1992qy, DeWolfe:2010he}.

\subsection{Critical exponent-\texorpdfstring{$\beta$}{beta} along first order line}\label{sec:beta}

For the first-order transition case, the true minimum of the free energy jumps from the lower to the upper branch at $T_c$ and $s$ is discontinuous (see the right plot of Fig.~\ref{s3}). The discontinuity of entropy density $s$ across the first-order line gives rise to the critical exponent $\beta$.
\begin{equation}\label{beta}
    \Delta s=s_>-s_<\sim (T_{\text{CEP}}-T)^\beta.
\end{equation}
At any generic point on the first-order line, $\Delta s$ is finite but reduces to zero when approaching the CEP along that line. The data is visualized using a log-log plot in Fig.~\ref{fig-beta}. The slope of a best fit line yields
\begin{equation}
    \beta = 0.322.
\end{equation}
This holographic outcome quantitatively agrees with experimental data and  the 3D Ising model~\cite{Goldenfeld:1992qy, DeWolfe:2010he}. Moreover, as we approach the critical endpoint, the entropy density at CEP, denoted as $s_{\text{CEP}}$, can be deduced from the converging values of $s_<$ and $s_>$. We then obtain
\begin{equation}\label{s-cep}
    \hat{s}_{\text{CEP}} =\frac{s_{\text{CEP}}}{T_{\text{CEP}}^3}= 0.8106\,,
\end{equation}
which will be used to compute the critical exponent $\delta$ along the critical isotherm.
\begin{figure}
  \centering
  \includegraphics[width=0.65\textwidth]{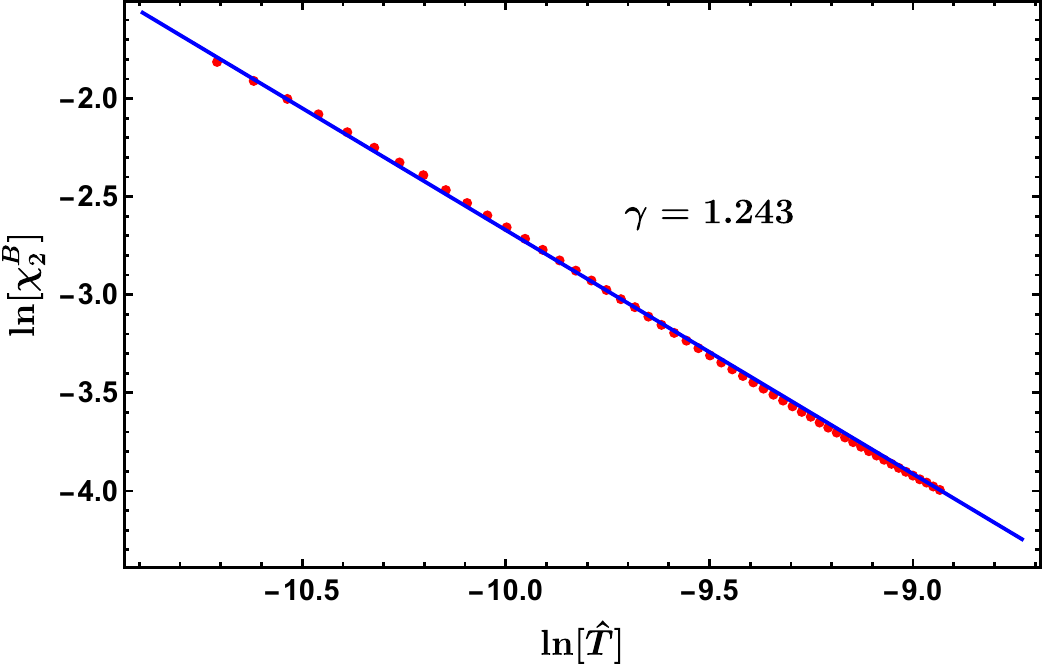}
  \caption{The baryon number susceptibility $\chi^B_2$ as a function of temperature $T$ as the CEP is approached on a log-log plot. We obtain the value of $\gamma$ from the slope, \emph{i.e.} $\gamma=1.243$. }\label{fig-gamma}
\end{figure}
%
\begin{figure}
  \centering
  \includegraphics[width=0.65\textwidth]{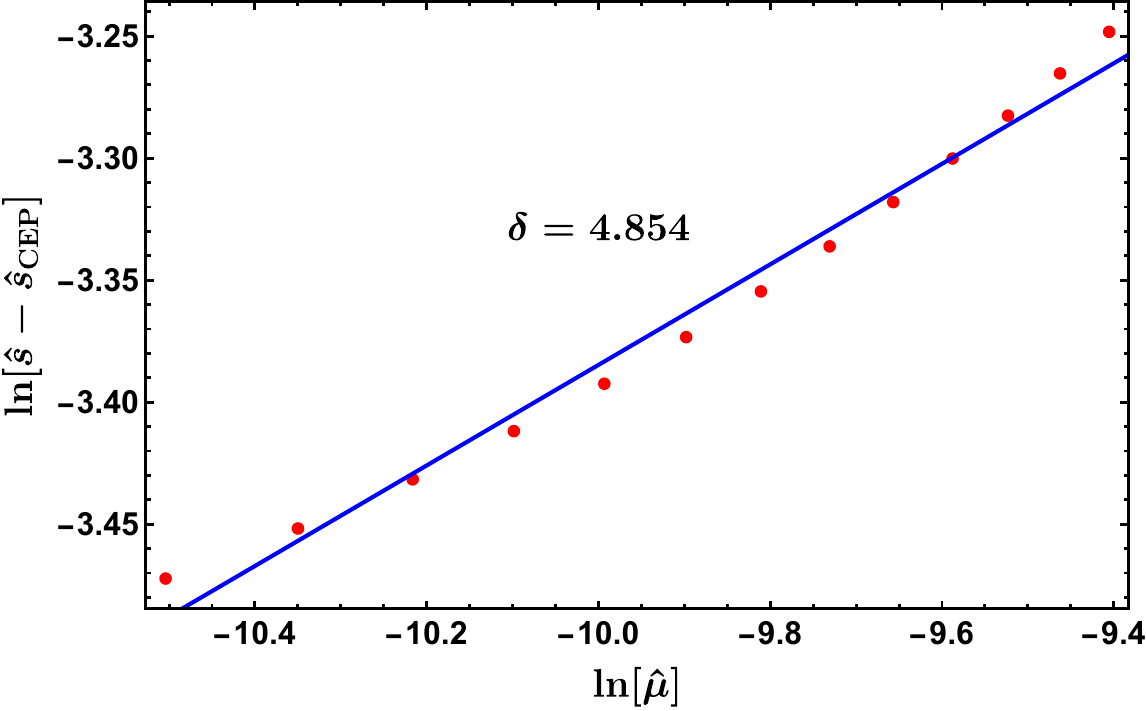}
  \caption{The trajectory of $s$ nearing $s_{\text{CEP}}$ is plotted as $\mu_B$ approaches $\mu_{\text{CEP}}$ on the critical isotherm. The derived slope from this plot gives us $\delta=4.854$.}\label{fig-delta}
\end{figure}
\subsection{Critical exponent-\texorpdfstring{$\gamma$}{gamma} along first order axis}\label{sec:gamma}

The exponent $\gamma$ is defined by the power law behavior of the baryon number susceptibility as the critical endpoint is approached along the tangent of the first-order line.
\begin{equation}\label{gamma}
    \chi^B_2=\frac{1}{T^2}\left( \frac{\partial n_B}{\partial \mu_B}\right)_T\sim |T-T_{\text{CEP}}|^{-\gamma}.
\end{equation}

Presenting the value of $\chi^B_2$ in a log-log plot with $\hat{T}=(T-T_{\text{CEP}})/T_{\text{CEP}}$, we find 
\begin{equation}
    \gamma =1.243.
\end{equation}
Once again, our holographic prediction is consistent with both the experimental measurement in fluids and the 3D Ising model~\cite{Goldenfeld:1992qy, DeWolfe:2010he}. 
\subsection{Critical exponent-\texorpdfstring{$\delta$}{delta} along critical isotherm}\label{sec:delta}

Now, let's calculate the last critical exponent $\delta$. The definition of $\delta$ is based on the relationship between $s-s_{\text{CEP}}$ and $\mu_B-\mu_{\text{CEP}}$ at the critical isotherm with $T=T_{\text{CEP}}$.
\begin{equation}
    s-s_{\text{CEP}}\sim|\mu_B-\mu_{\text{CEP}}|^{1/\delta}\,,
\end{equation}
where the value of $s_{\text{CEP}}$ has been given in Eq.~\eqref{s-cep}. Plotting $s-s_{\text{CEP}}$ in a log-log plot with $\hat{\mu}=\frac{\mu_B-\mu_{\text{CEP}}}{\mu_{\text{CEP}}}$, we obtain
\begin{equation}
    \delta = 4.854\,.
\end{equation}
The value of $\delta$ is once again in close alignment with experimental findings and the 3D Ising model~\cite{Goldenfeld:1992qy, DeWolfe:2010he}. 

The four critical exponents from our 2-flavor holographic model are summarized in Table.~\ref{tab-exp}. Following from the scaling behavior of the free energy at the critical endpoint, these thermodynamic exponents are not all independent. One should have the following scaling relations:
\begin{equation}\label{scalingrelation}
\alpha+2\beta+\gamma=2,\quad \alpha+\beta(1+\delta)=2\,.
\end{equation}
One can check that the values of our critical exponents quantitatively agree with the above scaling relations, providing a self-consistency check of our results.\footnote{In practice, it is more difficult to obtain $\alpha$ and $\delta$ as they require the location of CEP and numerical partial derivation with high precision. Nevertheless, one can use the scaling relations~\eqref{scalingrelation} to compute them since we know the values of $\beta$ and $\gamma$. The two approaches yield almost the same results.}

The general $O(N)$-symmetric universality classes are studied in \cite{Pelissetto:2000ek, Engels:2001bq, Engels:2003nq}. For a QCD system, the critical exponents strictly depend on the mass of quark flavors and the number of quark flavors. According to the ``Columbia plot" (refer to Figure 3 of~\cite{Ding:2015ona}), in the chiral limit of two quark flavors, where the masses of up (u) and down (d) quarks are zero and the mass of the strange (s) quark is very large (the left upper corner), the critical exponents belong to the $O(4)$ universality class~\cite{Engels:2001bq} ($\beta=0.38, \gamma=1.4668, \delta=4.86$). Conversely, when the masses of the u, d, and s quarks are relatively small (the left lower corner), the critical exponents follow the $Z(2)$ symmetric Ising universality class~\cite{Campostrini:2002cf, Ding:2015ona, Bazavov:2017xul} for which $\alpha=0.1096(5), \beta= 0.32653(10), \gamma=1.2373(2), \delta= 4.7893(8)$. In this work, we consider a finite mass quark system where the critical exponents are very close to the lattice results~\cite{Gupta:2014qka}, leading us to results that are closer to the $Z(2)$ universality class.

In Table.~\ref{tab-exp} we also compare our critical exponents with those from the experiments in non-QCD fluids, the full quantum 3D Ising model, the mean-field (van der Waals) theory, and the DGR model~\cite{Goldenfeld:1992qy, DeWolfe:2010he}. The results show that the critical exponents from $N_f=2$ holographic model closely align with experimental measurements in liquid/gas transition and the 3D Ising model's estimations. It suggests that the critical behavior of the CEP falls into the universality class of the 3D Ising model or the liquid/gas transition, and this result indirectly indicates that our holographic model surpasses the mean field level. Due to the finite mass effects of the u and d quarks, compared to the $O(4)$ universality class, they are closer to the $Z(2)$ universality class.
\begin{table}[ht!]
    \centering
    \setlength{\tabcolsep}{3.5mm}{
    \begin{tabular}{|c|c|c|c|c|c|}
     \cline{1-6}
      &  Experiment      &  3D Ising         & Mean field & DGR model & Ours   \\ \cline{1-6}
    $\alpha$ & 0.110-0.116 &  0.110(5)         &  0         &   0          & 0.113  \\ \cline{1-6}
    $\beta$  & 0.316-0.327 &  0.325$\pm$0.0015 & 1/2        & 0.482        & 0.322  \\ \cline{1-6}
    $\gamma$ &  1.23-1.25  &  1.2405$\pm$0.0015& 1          & 0.942        & 1.243  \\ \cline{1-6} 
    $\delta$ &  4.6-4.9    &  4.82(4)          & 3          & 3.035        & 4.854  \\ \cline{1-6} 
    \end{tabular}}
    \caption{Critical exponents from experiments in non-QCD fluids, the full quantum 3D Ising model, mean-field (van der Waals) theory, the DGR model and our 2-flavor holographic model.}
    \label{tab-exp}
\end{table}
\section{Conclusions}\label{sec:summary}
We have employed a holographic EMD theory to study the phase structure of $N_f=2$ QCD matter at finite temperature and baryon chemical potential, where all thermodynamic quantities are computed directly from the holographic renormalization. The model parameters are fixed completely by matching with the lattice QCD simulation at $\mu_B=0$ (see the EOS and second-order baryon susceptibility in Fig.~\ref{latt-EOS}). Moreover, the baryon number density $n_B$ versus $T$ at small $\mu_B$ also quantitatively agree with the lattice data. Notably, we have computed higher-order baryon number susceptibilities $\chi^B_n$ which show a rapid increase in their magnitudes near the pseudo-critical temperature and qualitatively agree with the available lattice data. We have used the Polyakov loop as an effective probe characterizing the phase transition. 

Through a thorough analysis of the behaviors of the free energy and the Polyakov loop, we have constructed the phase diagram in terms of $T$ and $\mu_B$. As visible from Fig.~\ref{PTD}, as $\mu_B$ increases, the crossover on the $T$-axis is sharpened into a first-order line at the critical endpoint. We have managed to give the exact location of the CEP, $(\mu_{\text{CEP}},\, T_{\text{CEP}})=(219\,\text{MeV}, \, 182\, \text{MeV})$, and the phase boundary for the first-order phase transition. To obtain the critical exponents associated with the CEP, we have systematically studied the approach of various thermodynamic quantities to criticality.
We have found that $\alpha=0.113,\,\beta=0.322, \,\gamma=1.243, \,\delta=4.854$, consistent with the scaling relations~\eqref{scalingrelation}. These critical exponents are in sharp contrast to mean-field theory, but they are quantitatively agree with with the experimental measurements in liquid/gas transition and the theoretical computation from 3D Ising model. Therefore, the critical behavior of the CEP should fall into the universality class of the 3D Ising model (or the liquid/gas transition).

We have limited to the EOS and critical phenomena in the present study, it will be interesting to consider the spectra and transport by considering the fluctuations around our hairy black hole backgrounds. One could also introduce the chiral symmetry in addition to the baryon number and compute the quark condensates. The generalization of our discussions to real-time dynamics far from equilibrium would be also very interesting . We hope to study these issues in the future.

\acknowledgments

We would like to thank Saumen Datta and Hai-cang Ren for helpful discussions. This work is supported in part by the National Key Research and Development Program of China under Contract No. 2022YFA1604900, the National Natural Science Foundation of China (NSFC) under Grant Nos.  11890711, 12075298, 11890710, 12275104, 11947233, 12075101, 12235016, and 12122513. S.H. is grateful for financial support from the Fundamental Research Funds for the Central Universities and the Max Planck Partner Group. 


\providecommand{\href}[2]{#2}\begingroup\raggedright\endgroup

\end{document}